# Mining Maximal Dynamic Spatial Co-Location Patterns

Xin Hu, Guoyin Wang, *Senior Member, IEEE*, and Jiangli Duan

*Abstract*—A spatial co-location pattern represents a subset of spatial features with instances that are prevalently located together in a geographic space. Although many algorithms for mining spatial co-location patterns have been proposed, the following selected problems remain: 1) these methods miss certain meaningful patterns (e.g., {*Ganoderma_lucidum$_{new}$*, *maple_tree$_{dead}$*} and {*water_hyacinth$_{new}$*(increase), *algae$_{dead}$*(decrease)}) and obtain the wrong conclusion if the instances of two or more features increase/decrease (i.e., new/dead) in the same/approximate proportion, which has no effect on the prevalent patterns; 2) because the number of prevalent spatial co-location patterns is quite large, the efficiency of existing methods is low in mining prevalent spatial co-location patterns. Therefore, we first propose the concept of the dynamic spatial co-location pattern that can reflect the dynamic relationships among spatial features. Second, we mine a small number of prevalent maximal dynamic spatial co-location patterns that can derive all prevalent dynamic spatial co-location patterns, which can improve the efficiency of obtaining all prevalent dynamic spatial co-location patterns. Third, we propose an algorithm for mining prevalent maximal dynamic spatial co-location patterns and two pruning strategies. Finally, the effectiveness and efficiency of the proposed method and the pruning strategies are verified by extensive experiments over real/synthetic datasets.

## I. Introduction

SPATIAL co-location pattern mining is an important component of association rule mining [1,2] in machine learning [3,4,5,6]. A spatial co-location pattern represents a subset of spatial features whose instances are prevalently located together in a geographic space. Mining of the spatial co-location pattern is significant. For example, if a city planner cannot find the prevalent pattern {*school, supermarket, restaurant*} near the "*school*", this indicates that we need to build a new "*supermarket*" or "*restaurant*" around the "*school*". Other application domains include public health [7], public transportation [8,9], environmental management [10], social media services [11,12], location services [13,14], and multimedia [15,16,17,18,19], among others.

This work is supported by the Startup Foundation for Introducing Talent of Yangtze Normal University (No. 0107/011160052), the National Key Research and Development Program of China (No. 2016YFB1000901) and the National Natural Science Foundation of China (NSFC, No.61936001, No. 61572091 and No. 61772096). (Corresponding author: Jiangli Duan).

X. Hu is with the College of Big Data and Intelligent Engineering, Yangtze Normal University, Chongqing 408100, China (huxin@yznu.edu.cn).

G. Wang and J. Duan are with the Chongqing Key Laboratory of Computational Intelligence, Chongqing University of Posts and Telecommunications, Chongqing 400065, China (wanggy@cqupt.edu.cn, jl_duan@126.com).

Although many methods for mining spatial co-location patterns exist, they cannot find the dynamic relationships among spatial features. On the one hand, the existing methods miss certain meaningful patterns. Case 1: "*Ganoderma_lucidum*" grows on the "*maple_tree*", which was dead. However, existing methods mine patterns from the set of coexisting plants such that the meaningful pattern {*Ganoderma_lucidum$_{new}$*, *maple_tree$_{dead}$*} was missed. Case 2: For mutually inhibitory features such as "*water_hyacinth*" and "*algae*", the instances of "*algae*" decrease with the increase in the instances of "water *hyacinth*" in the same zone. However, because the participation index is always unchanged for existing methods, these methods obtain the prevalent pattern {*water_hyacinth, algae*} regardless of the increase/decrease in instances of "*water_hyacinth*"/"*algae*". On the other hand, existing methods obtain the wrong conclusion that the instances of two or more features increase/decrease (i.e., new/dead) in the same/approximate proportion, which has no effect on prevalent patterns. Case 3: One application of the prevalent spatial co-location pattern judges whether the environment was polluted or not by comparing prevalent patterns at different time points. As shown in Fig. 1, the instances of two features were dead (black shadow) by an equivalent (or approximate) percentage because of environment disruption, but the existing methods determine that the environment has not been polluted because they obtain the same prevalent patterns with the same participation index at two-time points (i.e., $t_0$ and $t_1$). In conclusion, finding the dynamic relationships among spatial features (i.e., dynamic spatial co-location patterns) is a promising topic.

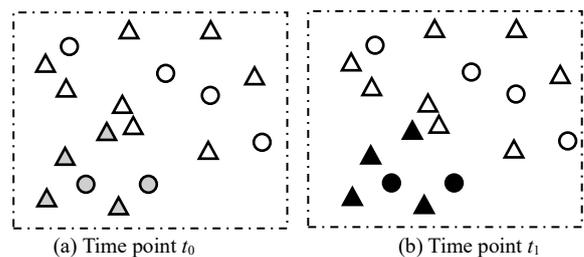

(a) Time point $t_0$    (b) Time point $t_1$
Fig. 1. Sample of case 3

In similar existing methods, the negative/sequential/strong-symbiotic patterns with the above cases appear to be similar, but essential differences exist between them. Both the dynamic pattern and the negative pattern [34,35] can mine mutual exclusion relationship, but the features in the negative pattern cannot coexist. Both the dynamic pattern and the sequential pattern [36,37] are all about time, but the latter is to look for prevalent subsequences from sequential databases. In contrast, the dynamic spatial co-location pattern represents a dynamic relationship between features, which exists in a symbiont circle. The strong-symbiotic pattern [38] belongs to a portion of the dynamic spatial co-location pattern, so it can only mine a portion of the dynamic spatial co-location patterns (i.e., {$A_{new}$, $B_{new}$}) and misses other dynamic spatial co-location patterns (i.e., {$A_{new}$, $B_{dead}$} and {$A_{dead}$, $B_{dead}$}).

Moreover, because the amount of spatial data is always quite large, the number of prevalent dynamic spatial co-location pattern is also large, and thus the efficiency of the existing methods is low in mining the prevalent spatial co-location patterns. It is necessary to find certain representative patterns that can derive all prevalent patterns and whose number is small. The prevalent maximal pattern is a compact representation of the prevalent pattern, and the number of prevalent maximal patterns is far less than the number of all prevalent patterns. Therefore, mining the prevalent maximal spatial co-location patterns that can derive all prevalent spatial co-location patterns is more efficient than mining all prevalent spatial co-location patterns using the existing methods. Although selected methods [39,40,41,42,43,44] can mine the prevalent maximal spatial co-location patterns, they still require large numbers of calculations and connections for table instances as well as general methods of mining prevalent spatial co-location patterns.

In summary, the existing methods cannot find the dynamic relationships among spatial features (i.e., dynamic spatial co-location pattern), and the efficiency of mining prevalent dynamic spatial co-location patterns by the existing methods is rather low. Therefore, in this paper, we propose a method for mining the prevalent maximal dynamic spatial co-location pattern and make the following contributions:

1. The existing methods cannot find the dynamic relationships among spatial features, and thus we propose the concept of the dynamic spatial co-location pattern (*Dc* for short) that can reflect the dynamic relationships among spatial features and can solve the problems in Case 1/Case 2/Case 3.
2. The prevalent maximal patterns can be used to derive all prevalent patterns, and the number of prevalent maximal patterns is far less than the number of all prevalent patterns. Therefore, we mine the prevalent maximal dynamic spatial co-location patterns rather than all prevalent dynamic spatial co-location patterns, which is more efficient than mining all prevalent dynamic spatial co-location patterns using the existing methods.
3. Because a large number of calculations and connections are necessary for table instances in the existing methods for mining maximal patterns, these methods have low efficiency. To improve the efficiency of mining maximal patterns, we propose an algorithm for mining the prevalent maximal dynamic spatial co-location patterns, in which the calculation and connection for table instances are turned into the calculation and connection of dynamic features whose number is far less than that of the instances. Moreover, we propose two pruning strategies to further improve the efficiency.
4. We verified the effectiveness of our algorithm (i.e., we can find the dynamic relationships among spatial features), the representativeness of the prevalent maximal *Dc*, the efficiency of our algorithm (i.e., comparison with the join-based method), and the efficiency of two pruning strategies over real/synthetic datasets.

II. RELATED WORK

Although many methods for mining spatial co-location pattern have been proposed, no method exists that can mine the dynamic spatial co-location patterns. S. Shekhar et al. [20,21] defined the spatial co-location pattern for the first time and proposed the join-based algorithm. Subsequently, certain methods focused on many other interesting research directions, such as high utility patterns [22,23,24], redundancy reduction [25], improved efficiency [26], causal rules [27], competitive pairs [28], fuzzy objects [29], uncertain data [30,31,32,33], etc. However, the existing methods miss certain meaningful patterns (e.g., {$Ganoderma\_lucidum_{new}$, $maple\_tree_{dead}$} and {$water\_hyacinth_{new}$(increase), $algae_{dead}$(decrease)}) and obtain the wrong conclusion that the instances of two or more features increase/decrease (i.e., new/dead) in the same/approximate proportion, which has no effect on the prevalent patterns.

The dynamic spatial co-location pattern in this paper might appear to a negative pattern [34,35], sequential pattern [36,37] or strong symbiotic pattern [38], but their essences are different. The features in the negative pattern [34,35] cannot coexist, whereas the features in the dynamic spatial co-location pattern must coexist. Sequential patterns [36,37] represent prevalent repeated paths between items, which exist in the form of a sequence, whereas the dynamic spatial co-location pattern represents a dynamic relationship between features which exist in a symbiont circle. In a strong symbiosis pattern [38], at least one feature benefits from the pattern, so it belongs to a portion of the dynamic spatial co-location pattern, and the method of mining strong symbiosis patterns can only mine a small portion of dynamic spatial co-location patterns (i.e., {$A_{new}$, $B_{new}$}), and other dynamic patterns (i.e., {$A_{new}$, $B_{dead}$} and {$A_{dead}$, $B_{dead}$}) cannot be mined. In conclusion, the methods for mining a negative pattern, sequential pattern or strong symbiotic pattern cannot mine the dynamic spatial co-location pattern in this paper.

TABLE I
CATEGORY OF METHODS

| Categories | Innovations | Literatures |
|---|---|---|
| Traditional co-location pattern | Origin | [20,21] |
| | high utility patterns | [22,23,24] |
| | redundancy reduction | [25] |
| | improved efficiency | [26] |
| | causal rules | [27] |
| | competitive pairs | [28] |
| | fuzzy objects | [29] |
| | uncertain data | [30,31,32,33] |
| Similar methods | Negative pattern | [34,35] |
| | Sequential pattern | [36,37] |
| | Strong-symbiotic | [38] |
| Related methods | maximal pattern | [39,40,41,42,43,44] |

Although certain methods can mine the prevalent maximal spatial co-location pattern, they still require a large number of calculations and connections for table instances as well as general methods of mining prevalent spatial co-location patterns. Wang et al. [39] proposed an order-clique-based approach for mining maximal co-location pattern, and based on this approach, Yao et al. [40,41] proposed an ordered-instance-clique approach. Dai et al. [42] used an index structure similar to four binary trees to mine the maximal spatial co-location patterns. Bao et al. [43] mined the top-*k* longer size maximal co-location patterns. Wang et al. [44] mined the maximal sub-prevalent co-location patterns, which introduced star participation instances to measure the prevalence of co-location patterns, i.e., spatially correlated instances that cannot form cliques were also properly considered. However, the above methods still require a large number of calculations and connections for table instances. In contrast, we propose an algorithm for mining prevalent maximal dynamic spatial co-location patterns, which is based on a degree-based approach for the maximum clique/maximal co-location patterns [45,46] and turns the calculation and connection for table instances into calculation and connection of dynamic features, the number of which is far less than that of instances.

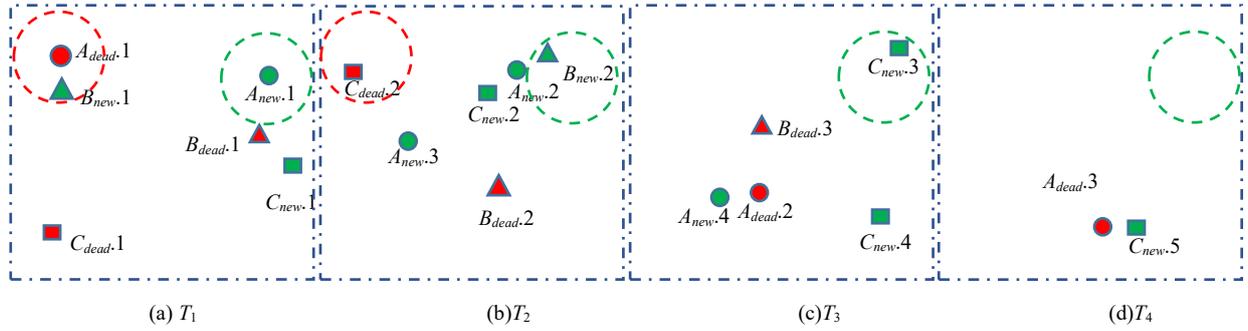

Fig. 2. Sample distribution datasets of new/dead instance

### III. BASIC DEFINITIONS

**Definition 1. (Dynamic Feature/Instance)** A dynamic feature represents the new/dead object in a certain area, denoted as $Df_{[i]}$, which is a new or dead object (e.g., $A_{new}$, $A_{dead}$ are two dynamic features in Fig. 2). A dynamic instance is an instance of a dynamic feature at a specific location, denoted as $Df_{[i]}.j$ (e.g., $A_{new}.1$ and $A_{new}.2$ are two instances of the dynamic feature $A_{new}$ in Fig. 2).

**Definition 2. (Dynamic Distance Threshold, $D_d$)** If the distance between two dynamic instances is less than $D_d$ as designated by experts, it is considered that the two dynamic instances have a relationship, and otherwise, they have no relationship.

**Definition 3. (Time Span and Time Span Constraint)** *Time span* is the time difference between two adjacent dynamic datasets and represents the time interval in which certain instances have changed, which is designated by experts. The *time span constraint* of a dynamic feature is the length of time in which the feature influences the surrounding dynamic features, denoted as $span (Df_{[i]})$. Furthermore, if the *time span constraint* of $Df_{[i]}$ is equal to $k$ time spans, it can be represented as $span (Df_{[i]}) = k$ (*time spans*).

The effect of a new object $Df_{[i]}$ on the surrounding dynamic features is the life cycle of $Df_{[i]}$. For example, we assume that the *time span* is 3 years and that the life cycle of $A_{new}$ is 75 years, and thus $span (A_{new}) = 25$(*time spans*).

The effect of a dead object $Df_{[i]}$ on the surrounding dynamic features is a one-*time span* because the dead object has little effect on the surrounding dynamic features, and the one-*time span* can sufficiently cover the time interval in which a dead object influences the surrounding instances. Thus $span (A_{dead}) = 1$(*time span*).

**Definition 4. (Dynamic Spatial Neighborhood Relationship, $D_R$)** For two dynamic instances, if the distance between them is less than $D_d$, which is designated by experts, and the time difference between them is less than the maximum of the *time span constraints* of all dynamic features, it is considered that the two dynamic instances satisfy the dynamic spatial neighborhood relationship $D_R$.

$D_R (A_{new}.1, B_{dead}.2) \Leftrightarrow$
$$distance (A_{new}.1, B_{dead}.2) \leq D_d$$
and
$$\triangle T (A_{new}.1, B_{dead}.2) < max (span (A_{new}), span (B_{dead}))$$

**Definition 5. (Dynamic Spatial Co-Location Pattern, $Dc$)** A dynamic spatial co-location pattern contains multiple new/dead features and can reflect the dynamic relationships among dynamic features, denoted as $Dc$. For example, $Dc_{[i]} = \{A_{new}, B_{dead}\}$ is a size-2 dynamic spatial co-location pattern.

**Definition 6. (Dynamic Row Instance and Dynamic Table Instance)** For a dynamic spatial co-location pattern $Dc$ and a set of dynamic instances $DI$, if a one-to-one match exists between each dynamic instance in $DI$ and each dynamic feature in $Dc$ and any two dynamic instances in $DI$ satisfy the dynamic spatial neighborhood relationship, we say that $DI$ is a dynamic row-instance of $Dc$, denoted as dynamic row-instance ($Dc$). The dynamic table-instance of $Dc$ consists of all distinct dynamic row-instances of $Dc$, denoted as dynamic table-instance ($Dc$).

**Example 1.** For $Dc_{[i]} = \{A_{new}, B_{new}\}$, if both $(A_{new}.1, B_{new}.2)$ and $(A_{new}.2, B_{new}.2)$ satisfy the dynamic spatial neighborhood relationship, they are the dynamic row-instance of $Dc_{[i]}$, and $\{\{A_{new}.1, B_{new}.2\}, \{A_{new}.2, B_{new}.2\}\}$ is the dynamic table-instance of $Dc_{[i]}$, denoted as dynamic table-instance($Dc_{[i]}$) = $\{\{A_{new}.1, B_{new}.2\}, \{A_{new}.2, B_{new}.2\}\}$.

**Definition 7. (Dynamic Participation Ratio ($DPR$)/ Index ($DPI$))** The dynamic participation ratio $DPR (Dc, Df_{[i]})$ of dynamic feature $Df_{[i]}$ in a size-$k$ dynamic spatial co-location pattern $Dc = \{Df_{[1]}, Df_{[2]} \ldots Df_{[k]}\}$ is defined as follows:

$$DPR(Dc, Df_{[i]}) = \frac{\pi_{Df_{[i]}}(dynamic\_table\_instance(Dc))}{dynamic\_table\_instance(\{Df_{[i]}\})},$$

where $\pi$ is the relational projection operation with a duplication elimination. $DPI(Dc)$ of $Dc$ is defined as shown:
$$DPI(Dc) = min_{i=1}^{k}\{DPR(Dc, Df_{[i]})\},$$

If $DPI(Dc)$ is greater than a given minimum prevalence threshold *min_prev* that is designated by experts and is used to judge whether the pattern occurs prevalently or not, we say that $Dc$ is a prevalent dynamic spatial co-location pattern.

**Example 2.** For $Dc_{[i]} = \{A_{new}, B_{new}\}$, dynamic table-instance($Dc_{[i]}$) = $\{\{A_{new}.1, B_{new}.2\}, \{A_{new}.2, B_{new}.2\}\}$. In $Dc_{[i]}$, the number of dynamic instances of $A_{new}$ and $B_{new}$ are 2 and 1, respectively. In contrast, from Fig. 2, the total number of dynamic instances of $A_{new}$ and $B_{new}$ are 4 and 2, respectively. Therefore, $DPR (Dc_{[i]}, A_{new}) = 2/4 = 0.5$, $DPR (Dc_{[i]}, B_{new}) = 1/2 = 0.5$ and thus $DPI(Dc_{[i]}) = min\{DPR(Dc_{[i]}, A_{new}), DPR(Dc_{[i]}, B_{new})\} = 0.5$. If *min_prev*=0.3, $Dc_{[i]} = \{A_{new}, B_{new}\}$ is a size-2 prevalent dynamic spatial co-location pattern.

**Definition 8. (Prevalent Maximal Dynamic Spatial Co-Location Pattern)** Given a prevalent dynamic spatial co-location pattern $Dc = \{Df_{[l]}, \ldots Df_{[v]}\}$, for any $Df_{[i]} \in Df$ and $Df_{[i]} \notin Dc$, if any $Dc \cup Df_{[i]}$ is not a prevalent dynamic spatial co-location pattern, then $Dc$ is a prevalent maximal dynamic spatial co-location pattern.

**Definition 9. (Dynamic Spatial Feature Clique, *Dfc*)** Given a dynamic spatial feature set $Dfc = \{Df_{[l]}, \ldots, Df_{[v]}\}$, if any size-2 pattern $Dc_{[i]} = \{Df_{[j]}, Df_{[k]}\}$ ($Df_{[j]}, Df_{[k]} \in Dfc$ and $j \neq k$) is prevalent, then $Dfc$ is a dynamic spatial feature clique.

**Definition 10. (Maximal Dynamic Spatial Feature Clique)** Given a dynamic spatial feature set $Dfc = \{Df_{[l]}, \ldots, Df_{[v]}\}$, for any $Df_{[i]} \in Df$ and $Df_{[i]} \notin Dfc$, if $Dfc \cup Df_{[i]}$ is not a dynamic spatial feature clique, then $Dfc$ is a maximal dynamic spatial feature clique.

## IV. MINING PREVALENT MAXIMAL *Dc* (ALGORITHM MDC)

We propose an algorithm for mining the prevalent maximal dynamic spatial co-location pattern (i.e., Algorithm MDC), which is divided into three sub algorithms (i.e., *Algorithm 1, Algorithm 2, Algorithm 3*). For convenience of description, the dynamic spatial co-location pattern is abbreviated as *Dc* in this paper.

Because a large number of calculations and connections exist for table instances in the existing methods, these methods have low efficiency. To improve the efficiency of mining patterns, after obtaining the size-2 prevalent dynamic spatial co-location patterns, we convert them to a dynamic feature graph *DG* by *Algorithm 1* such that the calculation and connection for the table instances are turned into the calculation and connection of the dynamic features. Subsequently, we obtain the set of maximal dynamic feature clique *Dfc* from the dynamic feature graph *DG* by *Algorithm 2*. Finally, each maximal dynamic feature clique *Dfc* as a candidate maximal *Dc* is verified by *Algorithm 3*, and thus we can obtain prevalent maximal *Dc*.

### A. First Sub-Algorithm

First, we generate the distribution dataset of dynamic instances (i.e., new/dead instances) from the distribution dataset of spatial instances at different time points (*step 1* and *Example 3*). Second, with $D_d$, *time span* and *time span constraint* (i.e., life cycle) of each dynamic feature, we can confirm whether any two dynamic instances have a dynamic spatial neighborhood relationship or not and obtain a set of all dynamic neighborhood relationships (*step 2* and *Example 4*). Second, we obtain the dynamic table-instance of all size-2 *Dc* by arranging all dynamic neighborhood relationships, and thus we obtain the size-2 prevalent *Dc* by definition 8 and *min_prev* (*step 3/step 4* and *Example 5*). Finally, we transform all size-2 prevalent *Dc* to dynamic feature graph *DG* (*step 5* and *Example 6*).

---

**Algorithm 1**: Generating Dynamic Feature Graph *DG*

**Input:** (1) $Df = \{Df_{[1]}, Df_{[2]} \ldots, Df_{[n]}\}$: a set of dynamic spatial features;(2) $St = \{St_{[1]}, St_{[2]} \ldots, St_{[n]}\}$:the distribution dataset of spatial instances at different time points; (3) $Lc = \{Lc_{[1]}, Lc_{[2]}, \ldots, Lc_{[n]}\}$:a set of life cycle of all dynamic features;(4)$D_d$:a dynamic distance threshold;(5)min_prev:a minimum DPI threshold.
**Output:** *DG*: dynamic feature graph.
**Variables:** $S_T$:the distribution dataset of dynamic instances at different time points.
1: $S_T$=Gen_dynamic_instance_distribution(*St*)
2: $\delta_{DR}$ = Gen_dynamic_neighborhood (*Df*, $S_T$, $D_d$, *Lc*, *time_span*)
3: $\delta_{dynamic\_table\_instance}$ = Gen_dynamic_table_instance (*Df*, $\delta_{DR}$)
4: $\delta_{dynamic\_size2}$=Gen_size2_prevalent_Dc (*Df*, $\delta_{dynamic\ table\ instance}$, min_prev)
5: *DG* = Gen_dynamic_feature_graph ($\delta_{dynamic\_size\ 2}$)

---

**Example 3.** Given the distribution dataset of spatial instances at different time points (i.e., $t_1, t_2, \ldots, t_n$), because this paper studies the dynamic relationship among features, we obtain *n*-1 dynamic datasets that contain only new/dead instances by comparing two datasets at $t_i$ and $t_{i+1}$. For instance, we can obtain a dynamic dataset in Fig. 2(a) by comparing the two datasets at $t_1$ and $t_2$. The new/dead (green/red in Fig. 2) categorizations of the same object are denoted by two dynamic features (i.e., $A_{new}/A_{dead}$), respectively. One instance of feature *A* exists in $t_1$ and disappears at $t_2$, and it is used as a dead instance in $T_1$, denoted by $A_{dead}.1$. Similarly, if one instance of feature *A* appears at $t_2$ for the first time, it is used as a new instance in $T_1$, denoted by $A_{new}.1$. As shown in Fig. 2, the distribution datasets of new/dead instances are obtained by comparing the datasets at 5 time points.

**Example 4.** Suppose $D_d = k$ and $span(A_{new})=3$(*time spans*) (the time difference between $T_1$ and $T_2$ is one *time span*). To obtain the neighborhood instances of $A_{new}.1$ in Fig. 2(a), we should confirm whether $A_{new}.1$ and all other dynamic instances in Fig. 2(a)(b)(c)(d) (according to definition 3 and $span(A_{new})=3$(*time spans*)) satisfy the dynamic spatial neighborhood relationship or not by definition 4. Subsequently, we determine that $B_{new}.2$ and $C_{new}.3$ are the neighborhood instances of $A_{new}.1$. Similarly, to obtain the neighborhood instances of $A_{dead}.1$ in Fig. 2(a), we should confirm whether $A_{dead}.1$ and all other dynamic instances in Fig. 2(a)(b) (according to definition 3, and thus $span(A_{dead})=1$ (*time span*)) satisfy the dynamic spatial neighborhood relationship or not by definition 4. Finally, we find that $B_{new}.1$ and $C_{dead}.2$ are the neighborhood instances of $A_{dead}.1$.

**Example 5.** By arranging all neighborhood dynamic instance pairs of $A_{new}$ and $B_{new}$ (i.e., $A_{new}.1$ and $A_{new}.2$ are the neighborhood dynamic instances of $B_{new}.2$), we obtain the dynamic table-instances $\{\{A_{new}.1, B_{new}.2\}, \{A_{new}.2, B_{new}.2\}\}$ of $Dc_{[i]} = \{A_{new}, B_{new}\}$. Suppose *min_prev*=0.3 according to definition 7, because $DPI(Dc_{[i]}) = 0.5 > 0.3$, and thus $Dc_{[i]} = \{A_{new}, B_{new}\}$ is a size-2 prevalent *Dc*.

**Example 6.** Suppose certain size-2 prevalent dynamic spatial co-location patterns exist such as $\{A_{new}, B_{new}\}$, $\{A_{new}, C_{new}\}$, $\{A_{dead}, B_{new}\}$, $\{A_{dead}, B_{dead}\}$, $\{A_{dead}, C_{dead}\}$ and $\{B_{new}, C_{dead}\}$. In Fig. 3, each dynamic feature in all size-2 prevalent *Dc* and each size-2 prevalent *Dc* are treated as a vertex and an edge, respectively. For instance, because $\{A_{new}, B_{new}\}$ is a size-2 prevalent *Dc*, we should connect $A_{new}$ and $B_{new}$. In contrast, $A_{new}$ does not connect to $C_{dead}$ because $\{A_{new}, C_{dead}\}$ is not a prevalent *Dc*. Finally, we obtain a dynamic feature graph *DG*, as shown in Fig. 3.

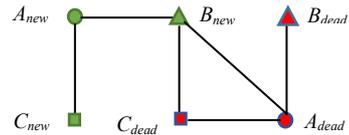

Fig. 3. Sample dynamic feature graph

### B. Second Sub-Algorithm

The maximal *Dfc* (dynamic feature clique) is treated as the candidate prevalent maximal *Dc* and can be obtained from *DG* by *Algorithm2,* which is proposed based on the degree-based approach for the maximum clique/maximal co-location patterns [45,46].

The core idea of *Algorithm2* is described as follows. First, the vertex with maximum degree in *DG* is selected as $V_{max}$, and other vertices are divided into two categories (adjacent and non-adjacent vertices). Second, $V_{max}$ is treated as a node in one candidate *Dfc*, and all of the adjacent vertices of $V_{max}$ form a subgraph (i.e., $sub\_DG_{[1]}$). This process is applied recursively for $sub\_DG_{[1]}$. Third, the non-adjacent vertices of $V_{max}$ are treated successively as $V_{max}'$, which is the same as $V_{max}$. For *i*-th $V_{max}'$, the adjacent vertices of $V_{max}'$ form a subgraph (i.e., $sub\_DG_{[2i]}$), and $sub\_DG_{[2i]}$ is subjected to a recursive process similar to that of $sub\_DG_{[1]}$.

In the recursive process, starting from the vertex with maximum degree ($V_{max}$) can accelerate the speed of finding all maximal *Dfc*. Because each vertex in the maximal *Dfc* must connect to all other vertices in the maximal *Dfc*, $V_{max}$ is always located in a maximal *Dfc* that contains many vertices. On the one hand, starting from $V_{max}$ can quickly reduce the edges in $sub\_DG_{[1]}$, and thus it accelerates the process of finding the maximal *Dfc* in $sub\_DG_{[1]}$. On the other hand, starting from $V_{max}$ always first finds the maximal *Dfc* that contains many vertices such that $sub\text{-}DG_{[2]}$ is far less than original *DG*, and therefore, it can accelerate the process of finding the maximal *Dfc* in $sub\_DG_{[2]}$.

---

**Algorithm 2**: Generating Maximal Dynamic Feature Clique

**Input:** *DG*: dynamic feature graph.
**Output:** $\delta_{maximal\_dynamic\_clique}$: the set of maximal *Dfc*.
**Variables:**(1)*Df*: a set of dynamic spatial features;(2)*can_ Dfc*: a set of candidate vertices in dynamic feature cliques;(3)$V_{max}$: a vertex with max degree;(4)*link_ Df*: a set of adjacent vertices of $V_{max}$;(5)*not_link_ Df*: a set of non-adjacent vertices of $V_{max}$;(6)$V_{max}'$: vertex in *not_link_Df* is regarded as $V_{max}'$ successively, which is the same as $V_{max}$;(7)*second_ Df*: a set of adjacent vertices of $V_{max}'$.
Clique (*Df*, *can_ Dfc*, *DG*)
1: $V_{max}$ = Get_maxdegree_ *Df* (*Df*, *DG*)
2: *link_ Df*, *not_link_ Df* =Get_link_or_unlink_$V_{max}$ (*Df*, *DG*)
3: If (Exist_side (*link_ Df*))
    Clique (*link_ Df*, *can_ Dfc* + $V_{max}$, *DG*)
   Else
    $\delta_{maximal\_dynamic\_clique}$ + (*can_ Dfc* + $V_{max}$ + *link_ Df*)
4: While (Exist_dynamic_feature (*not_link_ Df*, *DG*))
    $V_{max}'$ = Get_and_delete_first_ *Df* (*not_link_ Df*, *DG*)
    *second_ Df* = Get_link_$V_{max}'$ (*not_link_ Df*, *link_ Df*, *DG*)
    If (Exist_side(*second_ Df*, *DG*))
        Clique (*second_ Df*, *can_ Dfc* + $V_{max}'$, *DG*)
    Else
        $\delta_{maximal\_dynamic\_clique}$ + (*can_ Dfc* + $V_{max}'$ + *second_ Df*)

---

**Example 7.** For *DG* in Fig. 3, $V_{max\ [1]} = B_{new}$, and other vertices can be divided into adjacent/non-adjacent vertices. For the adjacent portion of $B_{new}$, $B_{new}$ is treated as one node in the candidate maximal *Dfc* (i.e., *can_Dfc*={$B_{new}$}), and the adjacent vertices form a subgraph (i.e., $sub\_DG_{[1]}$={$A_{new}$, $A_{dead}$, $C_{dead}$}). *Algorithm2* executes a recursion process for $sub\_DG_{[1]}$, $V_{max[11]}=A_{dead}$, and for the adjacent portion of $A_{dead}$, $A_{dead}$ is treated as one node in candidate maximal *Dfc* (i.e., *can_Dfc*={$B_{new}$, $A_{dead}$}, and the adjacent vertices form a subgraph (i.e., $sub\_DG_{[11]}$={$C_{dead}$}). Because no edge exists in $sub\_DG_{[11]}$, *Algorithm2* adds the remaining nodes to the candidate maximal *Dfc* and obtains the first maximal *Dfc* (i.e., $Dfc_{[1]}$={$B_{new}$, $A_{dead}$, $C_{dead}$}. For the non-adjacent portion of $A_{dead}$, the candidate maximal *Dfc* is {$B_{new}$}, and no edge exists in $sub\_DG_{[12]}$={$A_{new}$}. Similarly, we obtain the second maximal *Dfc* (i.e., $Dfc_{[2]}$={$B_{new}$, $A_{new}$}). For the non-adjacent vertices (i.e., $C_{new}$ and $B_{dead}$) of $B_{new}$, the candidate maximal *Dfc* is {}, $C_{new}$ and $B_{dead}$ are successively treated as $V_{max}'$, and their adjacent vertices form subgraphs $sub\_DG_{[21]}$={$A_{new}$} and $sub\_DG_{[22]}$={$A_{dead}$}, respectively. Thus, we can obtain $Dfc_{[3]}$={$C_{new}$, $A_{new}$} and $Dfc_{[4]}$={$B_{dead}$, $A_{dead}$}. Finally, we can obtain all maximal *Dfc* (i.e., {$B_{new}$, $A_{dead}$, $C_{dead}$}, {$B_{new}$, $A_{new}$}, {$C_{new}$, $A_{new}$} and {$B_{dead}$, $A_{dead}$}).

### C. Third Sub-Algorithm

Each maximal *Dfc* is treated as a candidate maximal *Dc* (i.e., $Dc_{[i]}$); therefore, we need to verify whether it is prevalent or not. First, we obtain the dynamic table instance of the candidate maximal $Dc_{[i]}$, which can be obtained by the dynamic table-instance of the size-2 prevalent *Dc* (*Example 8*). Second, we calculate the *DPI* of the candidate maximal $Dc_{[i]}$ and compare it with *min_prev*. Third, if the $Dc_{[i]}$ is prevalent, we add it to the set of prevalent maximal *Dc*; otherwise, it is decomposed into a size-*k*-1 pattern ($Dc_{[i]}$ is a size-*k* pattern) (*Example 9*), and each size-*k*-1 pattern is treated as a candidate maximal *Dc* and passes through a verification process (except the one that already exists in the set of prevalent maximal *Dc*).

---

**Algorithm 3:** Verifying Prevalent Maximal Spatial Co-Location Pattern
**Input:** (1)$\delta_{maximal\_Dfc}$: the set of maximal *Dfc* which is regarded as candidate prevalent maximal *Dc*;(2)*min_prev*: minimum prevalent threshold;(3)$\delta_{size\ 2\_prevalent}$: size-2 prevalent *Dc*.
**Output:** $\delta_{prevalent\_dynamic\_maximal\_Dc}$:the set of prevalent *Dc*.
**Variables:** (1)*clique*: one maximal *Dfc*;(2)$Df_{[n]}$: a set of dynamic features in *clique*;(3)$\delta_{clique\_size\ 2\_prevalent}$: a set of dynamic table instances of size-2 prevalent *Dc* which contains $Df_{[0]}$;(4)*common_instance*: common instances of $Df_{[0]}$ in size-2 prevalent *Dc* which contains $Df_{[0]}$;(5)$\delta_{cci}$: validated dynamic table instance of *Dfc*;(6)$\delta_{size\_k-1}$: a set of size-*k*-1 subpatterns of *Dfc*.
While (not_empty($\delta_{maximal\_Dfc}$))
1: *clique* = Get_one_clique($\delta_{maximal\_Dfc}$)
2: $Df_{[n]}$ = Get_dynamic_feature(*clique*)
3: for (*i* = 1; *i* < *n*; *i* ++)
    $\delta_{clique\_size2\_prevalent}$=Get_dynamic_table_instance($Df_{[0]}$,$Df_{[i]}$,$\delta_{size2\_prevalent}$)
4: *common_instance* = Get_common_instance_$Df_{[0]}$ ($\delta_{clique\_size2\_prevalent}$)
5: $\delta_{cci}$=Get_clique_dynamic_instance(*common_instance*,$\delta_{clique\_size2\_prevalent}$)
6: for (*i* = 1; *i* < *n* - 1; *i* ++)
    for (*j* = *i* + 1; *j* < *n*; *j* ++)
        $\delta_{cci}$ = Verifying($\delta_{cci}$, $\delta_{clique\_dynamic\_size\ 2}$, $Df_{[i]}$, $Df_{[j]}$)
7: If (prevalently($\delta_{cci}$))
        $\delta_{prevalent\_dynamic\_maximal\_Dc}$ = $\delta_{prevalent\_dynamic\_maximal\_Dc}$ + *clique*
8: Else if (size (*clique*) > 3)
        $\delta_{size\_k-1}$ = Split (*clique*)
        $\delta_{size\_k-1}$=Non_prevalent($\delta_{size\_k-1}$,$\delta_{prevalent\_dynamic\_maximal\_Dc}$,$\delta_{maximal\_Dfc}$)
        $\delta_{maximal\_Dfc}$ =$\delta_{maximal\_Dfc}$ + $\delta_{size\_k-1}$

---

**Example 8.** Given the maximal *Dfc*, which is also a candidate maximal *Dc* (i.e., $Dc_{[i]}$={$A_{dead}$, $B_{new}$, $C_{dead}$}), we first obtain the common instances from the dynamic table instances of selected size-2 prevalent subpatterns (i.e., $Dc_{[i1]}$={$A_{dead}$, $B_{new}$} and $Dc_{[i2]}$={$A_{dead}$, $C_{dead}$}). For example, from the dynamic table-instance($Dc_{[i1]}$)={{$A_{dead}$.1, $B_{new}$.1}, {$A_{dead}$.1, $B_{new}$.2}, {$A_{dead}$.2, $B_{new}$.1}} and the dynamic table-instance($Dc_{[i2]}$)={{$A_{dead}$.1, $C_{dead}$.2}}, we can obtain the common instances (i.e., $A_{dead}$.1) of $Dc_{[i1]}$ and $Dc_{[i2]}$. Second, with the common dynamic instances, we select the dynamic row-instances of the size-2 prevalent subpatterns to construct the candidate dynamic row-instances of $Dc_{[i]}$. For example, with $A_{dead}$.1, we select the dynamic row-instances set {{$A_{dead}$.1, $B_{new}$.1}, {$A_{dead}$.1, $B_{new}$.2}} and {{$A_{dead}$.1, $C_{dead}$.2}} and subsequently construct the dynamic row-instances of $Dc_{[i]}$ (i.e., {$A_{dead}$.1, $B_{new}$.1, $C_{dead}$.2} and {$A_{dead}$.1, $B_{new}$.2, $C_{dead}$.2}. Third, we verify the dynamic row-instances by the other size-2 prevalent subpattern (i.e., $Dc_{[i3]}$={ $B_{new}$, $C_{dead}$}. For example, because only the dynamic row-instance {$B_{new}$.1, $C_{dead}$.2} exists in the dynamic table-instance of $Dc_{[i3]}$ while another dynamic row-instance {$B_{new}$.2, $C_{dead}$.2} does not exist, only {$A_{dead}$.1, $B_{new}$.1, $C_{dead}$.2} is a real dynamic row-instance of $Dc_{[i]}$, and thus we obtain the dynamic table-instance($Dc_{[i]}$)={{$A_{dead}$.1, $B_{new}$.1, $C_{dead}$.2}}.

**Example 9**. Supposing that a size-4 candidate *Dc* (i.e., $Dc_{[i]}$={$A_{dead}$,$B_{new}$,$C_{dead}$,$D_{new}$}) is not prevalent, we divide the *Dc* into size-3 candidate *Dc* (i.e., {$A_{dead}$, $B_{new}$, $C_{dead}$}, {$A_{dead}$, $B_{new}$, $D_{new}$}, {$A_{dead}$, $C_{dead}$, $D_{new}$} and {$B_{new}$, $C_{dead}$, $D_{new}$}). If candidate pattern {$A_{dead}$, $B_{new}$, $C_{dead}$} already exists in the set of prevalent maximal *Dc*, then it is deleted from the set of subpatterns. We add the size-3 candidate patterns {$A_{dead}$, $B_{new}$, $D_{new}$}, {$A_{dead}$, $C_{dead}$, $D_{new}$} and {$B_{new}$, $C_{dead}$, $D_{new}$}} to the set of candidate *Dc*.

## V. PRUNING STRATEGIES

**Theorem 1**. For a dynamic feature $Df_{[j]}$ in the candidate prevalent maximal $Dc_{[i]}$, if $DPR(Dc_{[i]}, Df_{[j]})$<*min_prev*, then $Dc_{[i]}$ is non-prevalent.

*Proof*. From definition 7, $DPI(Dc_{[i]})$ is the minimum value among dynamic participation ratios of all dynamic features in $Dc_{[i]}$. Therefore, if $DPR(Dc_{[i]}, Df_{[j]})$ <*min_prev*, then $DPI(Dc_{[i]})\le DPR(Dc_{[i]}, Df_{[j]}) \le$ *min_prev*, namely, $Dc_{[i]}$ is non-prevalent.

**Pruning Strategy 1**. During verification of a candidate prevalent maximal *Dc*, if the dynamic participation ratio of any dynamic feature is smaller than *min_prev*, then we can stop verification and confirm that the candidate maximal *Dc* is nonprevalent.

**Example 10**. During verification of the size-4 candidate prevalent maximal *Dc* (i.e., $Dc_{[i]}$={$A_{dead}$, $B_{new}$, $C_{dead}$, $D_{new}$}), we can calculate the dynamic participation ratio of any dynamic feature in the process of obtaining the common instances of $A_{dead}$ and verifying the dynamic row-instance of $Dc_{[i]}$. If $DPR(Dc_{[i]}, A_{dead})$ <*min_prev*, then we can stop verification and confirm that $Dc_{[i]}$ is non-prevalent.

**Theorem 2**. Given a candidate prevalent maximal $Dc_{[i]}$ and its superpattern $Dc_{[i]}'$ (i.e., $Dc_{[i]} \subseteq Dc_{[i]}'$ and $Dc_{[i]} \ne Dc_{[i]}'$), if $Dc_{[i]}$ is non-prevalent, then $Dc_{[i]}'$ is also non-prevalent.

*Proof*. Pattern $Dc_{[i]}$ is non-prevalent, and we know that a dynamic feature $Df_{[j]} \in Dc_{[i]}$ exists and $DPR(Dc_{[i]},Df_{[j]})<$ *min_prev*. Therefore, for its superpattern $Dc_{[i]}'$ (i.e., $Dc_{[i]} \subseteq Dc_{[i]}'$ and $Dc_{[i]} \ne Dc_{[i]}'$), the existence of inheritance leads to $DPR(Dc_{[i]}', Df_{[j]}) \le DPR(Dc_{[i]},Df_{[j]})<$*min_prev*, and thus $Dc_{[i]}'$ is non-prevalent.

**Pruning Strategy 2**. If multiple candidate prevalent maximal dynamic spatial co-location patterns have a common subpattern, we can first verify their common subpattern, and if the common subpattern is non-prevalent, all candidate prevalent maximal spatial co-location patterns are non-prevalent.

**Example 11**. Given a candidate prevalent maximal dynamic spatial co-location patterns $Dc_{[i]}$={$A_{dead}$, $B_{new}$, $C_{dead}$, $D_{new}$} and $Dc_{[j]}$={$A_{dead}$, $B_{new}$, $C_{dead}$, $E_{dead}$}, we can first verify the common subpattern $Dc_{[ij]}$={$A_{dead}$, $B_{new}$, $C_{dead}$}, and if $Dc_{[ij]}$ is non-prevalent, then both $Dc_{[i]}$ and $Dc_{[j]}$ are non-prevalent. If $Dc_{[ij]}$ is prevalent, then we verify the other portions of $Dc_{[i]}$ and $Dc_{[j]}$.

## VI. COMPLEXITY ANALYSIS

To analyze the complexity, the upper limits of certain parameters will need to be determined. For the original datasets (i.e., *n* time points), the number of instances at each time point is no more than *I*. For the dynamic datasets (i.e., *n*-1 time points), the number of dynamic instances in each dynamic dataset is $I'$, and there are $2*F$ dynamic features (new/dead). For any dynamic feature, its dynamic instances and *time span constraint* are no more than *i* and *n*, respectively.

### A. Time Complexity

In *Algorithm1*, from step 1 to step 5, the corresponding complexities are $O(I*I*(n-1))$, $O(I'^2*(n-1)^2)$, $O(I'^2*(n-1)^2/2)$, $O(F*F)$ and $O(F*F)$, and thus the time complexity of *Algorithm1* is $O(I^2*n)+O(I'^2*n^2)$, where *F* is much less than *I*, so selected portions have been omitted. In *Algorithm2*, for any dynamic feature $Df_{[j]}$, the number of dynamic feature cliques that are related to $Df_{[j]}$ is no more than *F*, and there are $2*F$ dynamic features. Therefore, the number of dynamic feature cliques is no more than $F*2*F$. Moreover, from the literature [45], the time complexity of obtaining a maximum clique is $O(1.442^F)$, and thus the time complexity of *Algorithm2* is $O(1.442^F*F^2)$. In *Algorithm3*, the number of dynamic feature cliques is $F*2*F$ (from *Algorithm2*), any dynamic feature clique is decomposed once (on average), and therefore, its number is no more than $F*2*F*F$ after decomposition. Moreover, for one dynamic feature clique, the time complexity of verification is $O(F^2)$, so the time complexity of *Algorithm3* is $O(F^5)$. Therefore, the time complexity of Algorithm MDC is $O(I^2*n)+O(I'^2*n^2)+O(1.442^F*F^2)+O(F^5)$.

### B. Space Complexity

The space complexity values of the storing instances, dynamic features, dynamic instances, adjacent instance set, dynamic table-instances, dynamic feature graph, dynamic feature cliques, common code and decomposed maximal cliques are $O(I*n)$, $O(2*F)$, $O(I'*(n-1))$, $O(I'^2*(n-1)^2/2)$, $O(I'^2*(n-1)^2/2)$, $O(F*F)$, $O(F*2*F)$, $O(i)$ and $O(F)$, respectively. Moreover, during searching of the dynamic feature clique, the space complexity is $O(F^2)$. Therefore, the space complexity of Algorithm MDC is $O(I*n)+O(I'^2*n^2)+O(F*2*F)$, where $I'$, *I* and *F* are much less than *I*, $I'*n$ and *I*, respectively, and thus certain portions have been omitted.

## VII. EXPERIMENTAL EVALUATION

Various experiments over both real and synthetic datasets were conducted to verify the effectiveness of Algorithm MDC (i.e., we can find the dynamic relationships among spatial features), the representativeness of the prevalent maximal *Dc*, the efficiency of Algorithm MDC (i.e., comparison with the join-based algorithm), and the efficiency of the two pruning strategies.

### A. Experiments on Real Dataset

In this section, we verify the effectiveness of the Algorithm MDC, namely, whether the Algorithm MDC can mine the dynamic relationships among spatial features (i.e., *Dc*) from the real datasets.

The real dataset is sourced from the Wuhua district of Kunming, Yunnan province, China, in the most recent 30 years. Specifically, *Df*= {"*School*", "*Park*", "*Hospital*", "*Hotel*", "*Supermarket*", "*KTV*", "*Bank*"}, where "*Bank*" includes bank business halls and ATMs, and "*Hospital*" includes clinics, pharmacies, etc. The *life cycle* of all dynamic feature is {30,30,15,9,6,6,6}, the number of new/dead instances is approximately 1500, and the *time span* is 3 years. If $D_d$ is 1 km and *min_prev* is 0.4, we can obtain all prevalent *Dc* as shown in Table II, which can be derived from the prevalent maximal *Dc*.

TABLE II
PREVALENT DYNAMIC SPATIAL CO-LOCATION PATTERNS

| | $A_{new}, B_{new}$ | $A_{dead}, B_{dead}$ | $A_{new}, B_{dead}$ |
|---|---|---|---|
| size-2 | $\{school_{new}, supermarket_{new}\}$ | $\{school_{dead}, bank_{dead}\}$ | $\{hotel_{dead}, supermarket_{new}\}$ |
| | $\{school_{new}, bank_{new}\}$ | $\{hospital_{dead}, supermarket_{dead}\}$ | $\{supermarket_{new}, KTV_{dead}\}$ |
| | $\{school_{new}, KTV_{new}\}$ | $\{hotel_{dead}, bank_{dead}\}$ | $\{hospital_{new}, KTV_{dead}\}$ |
| | $\{KTV_{new}, bank_{new}\}$ | $\{hotel_{dead}, KTV_{dead}\}$ | $\{school_{new}, KTV_{dead}\}$ |
| | $\{hospital_{new}, bank_{new}\}$ | $\{KTV_{dead}, bank_{dead}\}$ | |
| | $\{hospital_{new}, supermarket_{new}\}$ | | |
| | $\{hotel_{new}, bank_{new}\}$ | | |
| | $\{hotel_{new}, KTV_{new}\}$ | | |
| | $\{part_{new}, hotel_{new}\}$ | | |
| size-3 | $\{hotel_{new}, KTV_{new}, bank_{new}\}$ | $\{hotel_{dead}, KTV_{dead}, bank_{dead}\}$ | $\{hotel_{dead}, supermarket_{new}, KTV_{dead}\}$ |

Traditional methods can obtain the prevalent pattern {"*School*", "*Park*", "*Hospital*", "*Hotel*", "*Supermarket*", "*KTV*", "*Bank*"} at each time point (i.e., $t_0, t_1, t_2, \ldots, t_n$) when $D_d$=1 km and $min\_prev$=0.4, namely, all features are always coexistent at each time point, and the result is not meaningful. In contrast, from the experimental results of Algorithm MDC in Table II, we can obtain the following meaningful information:

1) The instances of "*Bank*" increase (or decrease) with the increase (or decrease) of the instances of "*School*", "*Hotel*", "*Hospital*" and "*KTV*", which means that Algorithm MDC can find the dynamic relationships among spatial features such as $\{water\_hyacinth_{new}(\text{increase}), algae_{dead}(\text{decrease})\}$.
2) Life service (e.g., "*Hospital*" and "*Supermarket*") has a mutual exclusion relationship with entertainment (e.g., "*KTV*"), namely, the instances of "*Hospital*" and "*Supermarket*" increase (or decrease) with the decrease (or increase) of the instances of "*KTV*", which represents the adjustment of an urban regional structure. This result means that Algorithm MDC can find the dynamic relationships among spatial features such as $\{Ganoderma\_lucidum_{new}, maple\_tree_{dead}\}$.
3) The instances of "*Hotel*", "*KTV*" and "*Bank*" always appear/disappear simultaneously, which indicates that they have strong symbiotic relationships, and they reflect the economic prosperity/recession in this region because they denote the level of regional economic development. This result means that Algorithm MDC can effectively avoid the wrong conclusion that the instances of two or more features increase/decrease (i.e., new/dead) in the same/approximate proportion, which has no effect on prevalent patterns.

In conclusion, Algorithm MDC can obtain certain meaningful patterns such as $\{water\_hyacinth_{new}(\text{increase}), algae_{dead}(\text{decrease})\}$ and $\{Ganoderma\_lucidum_{new}, maple\_tree_{dead}\}$, and avoid the wrong conclusion that the instances of two or more features increase/decrease (i.e., new/dead) in the same/approximate proportion, which has no effect on prevalent patterns."

### B. Experiments on Synthetic Datasets

In this section, we examine the representation of the prevalent maximal *Dc* for all prevalent *Dc*, the efficiency of Algorithm MDC, and the performance of the pruning strategies.

We randomly generate synthetic datasets, where the *time span* is 3, there are 11 time points, and the distribution area of spatial instances is 1000*1000. By default, the number of dynamic instances and dynamic features are 10000 and 10, respectively; the *life cycle* of all dynamic features is {9,3,30,15,27,24,30,3,24,18}; and $min\_prev$ and $D_d$ are 0.1 and 35, respectively.

#### 1) Change Trend of Maximal Dc and Dc

We analyze the change trend of the number of size-*k* ($k \in [1,10]$) patterns (i.e., the pattern of each size) and the sum of the number of patterns from size-1 to size-*k* (i.e., sum of patterns), as shown in Fig. 4.

Furthermore, the change trend of the number of maximal *Dc* and that of *Dc* are approximate to the blue line (the pattern of each size) and red line (i.e., sum of patterns), respectively. Suppose both of the size-*k* patterns and its low-size patterns are prevalent, and the size-*k*+1 patterns are non-prevalent. On the one hand, the subsets of the size-*k* patterns are prevalent and its superset are non-prevalent at this time, and by the definition of the maximal *Dc*, the size-*k* patterns are maximal prevalent patterns, which leads to the observation that the change trend of the number of prevalent maximal *Dc* is approximate to that of the size-*k* patterns (i.e., the pattern of each size (blue line)). On the other hand, all prevalent patterns include patterns from size-1 to size-*k*, which leads to the observation that the change trend of the number of prevalent *Dc* is approximate to the red line (i.e., sum of patterns). Finally, although the analysis of change trend starts from the supposed premise, its real change trend is actually approximate to the lines shown in Fig. 4.

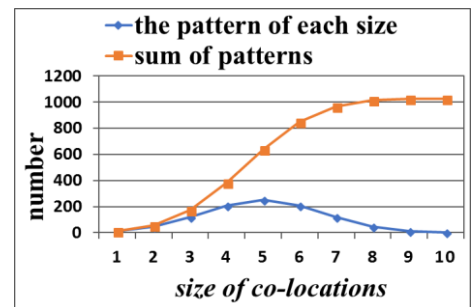

Fig. 4. Change trend of pattern with different size

In Fig. 4, the blue line first increases and subsequently decreases, which represents the change trend of the number of prevalent maximal *Dc* mined by our method. The red line continues to increase, which represents the change trend of the number of prevalent *Dc* mined by traditional methods. First, with the increase of size-*k*, the gap between the number of prevalent maximal *Dc* and that of prevalent *Dc* becomes larger, which means that mining the prevalent maximal *Dc* is more efficient than mining the prevalent *Dc*. Moreover, with the increase of size-*k*, the number of prevalent *Dc* significantly increases, which leads to the observation that the execution time of the traditional method is unacceptable, and thus we only display the complete trend (i.e., the prevalent maximal *Dc* first increases and subsequently decreases) in Fig. 5(a), which is treated as a sample example and partial trend (i.e., the prevalent maximal *Dc* increases) in other comparison experiments.

*2) Representativeness of Prevalent Maximal Dc*

We compare the number of prevalent maximal *Dc* with the number of prevalent *Dc* over the change in *number of dynamic instances*, $D_d$, *min_prev*, and *number of dynamic features,* as shown in Table III, Table IV, Table V and Table VI (Comparisons in Fig. 5 is more intuitive). With the increase in *number of dynamic instances*, the numbers of prevalent maximal *Dc* and prevalent *Dc* increase, which lead to the increase of size-*k* of the prevalent maximal *Dc* and prevalent *Dc* such that the number of prevalent maximal *Dc* first increases and subsequently decreases and that of the prevalent *Dc* continues to increase, as shown in Fig. 5(a) (or Table III) and Fig. 4.

TABLE III
REPRESENTATIVENESS OF VARIED NUMBER OF DYNAMIC INSTANCES

| dynamic instances | 5k | 6k | 7k | 8k | 9k | 10k | 11l |
|---|---|---|---|---|---|---|---|
| prevalent maximal *Dc* | 99 | 131 | 154 | 191 | 162 | 142 | 99 |
| prevalent *Dc* | 193 | 270 | 414 | 580 | 670 | 720 | 870 |

TABLE IV
REPRESENTATIVENESS OF VARIED $D_d$

| $D_d$ | 15 | 20 | 25 | 30 | 35 |
|---|---|---|---|---|---|
| prevalent maximal *Dc* | 44 | 73 | 115 | 153 | 170 |
| prevalent *Dc* | 44 | 101 | 214 | 376 | 583 |

TABLE V
REPRESENTATIVENESS OF VARIED *min_prev*

| *min_prev* | 0.25 | 0.20 | 0.15 | 0.10 | 0.05 |
|---|---|---|---|---|---|
| prevalent maximal *Dc* | 93 | 104 | 116 | 126 | 139 |
| prevalent *Dc* | 306 | 376 | 460 | 583 | 767 |

TABLE VI
REPRESENTATIVENESS OF VARIED NUMBER OF DYNAMIC FEATURES

| dynamic features | 9 | 10 | 11 | 12 | 13 |
|---|---|---|---|---|---|
| prevalent maximal *Dc* | 67 | 152 | 316 | 658 | 1096 |
| prevalent *Dc* | 416 | 752 | 1070 | 2059 | 3349 |

Because the execution time of the traditional method is unacceptable, we only display a partial trend (i.e., the prevalent maximal *Dc* increases) in other comparison experiments (more detailed information is given in the previous section). From Fig. 5(b)(c)(d) (or Table IV, Table V and Table VI), on the one hand, the number of prevalent maximal *Dc* is far less than the number of all prevalent *Dc*, which means that mining the prevalent maximal *Dc* by our method is more efficient than mining the prevalent *Dc* by traditional methods. On the other hand, with the change of $D_d$, *min_prev* and *number of dynamic features*, the gap between the number of prevalent maximal *Dc* and that of prevalent *Dc* becomes larger, and the difference in efficiency between mining the prevalent maximal *Dc* by our method and mining the prevalent *Dc* by the traditional method is increasingly obvious.

*3) Efficiency of Algorithm MDC*

We compare the running times for mining the prevalent *Dc* by Algorithm MDC and the join-based algorithm [15,16] over the change in *number of dynamic instances*, $D_d$, *min_prev* and *number of dynamic features,* as shown in Fig. 6. Because the number of prevalent *Dc* is small on the original parameters, the advantage of Algorithm MDC is not obvious compared with traditional method. Moreover, with the change in these parameters, the number of prevalent *Dc* increases, and the difference in efficiency between mining the prevalent maximal *Dc* by our method and mining the prevalent *Dc* by the traditional method is becomes more obvious.

*4) Performance of Pruning Strategies*

We compare the efficiency before and after pruning via pruning strategies 1 and 2, which can effectively accelerate the process of mining the prevalent maximal *Dc* by Algorithm MDC, as shown in Fig. 7.

VIII. CONCLUSION

Because the existing methods cannot mine the dynamic relationships among spatial features, and the number of prevalent patterns is too large, this paper proposes 1) the definition of a dynamic spatial co-location pattern, 2) the idea of mining the prevalent maximal patterns instead of all prevalent patterns (the former can be used to derive the latter), and 3) an algorithm for mining the maximal dynamic spatial co-location patterns (i.e., Algorithm MDC) based on the maximal dynamic feature clique.

Mining the dynamic spatial co-location pattern can remedy the defects of the existing methods. The existing methods miss certain meaningful patterns, such as {*Ganoderma_lucidum$_{new}$*, *maple_tree$_{dead}$*} and {*water_hyacinth$_{new}$*(increase), *algae$_{dead}$*(decrease)}, and obtain the wrong conclusion that the instances of two or more features increase/decrease (i.e., new/dead) in the same/approximate proportion, which has no effect on prevalent patterns. Therefore, we propose the dynamic spatial co-location pattern *Dc*, which can reflect the dynamic relationships among spatial features similar to the above three types of dynamic changes.

Compared with mining the prevalent *Dc*, mining the prevalent maximal *Dc* that can derive all prevalent *Dc* is more efficient. The number of prevalent patterns is large, which makes the efficiency of the existing methods low. Therefore, we introduce the prevalent maximal pattern into the process of mining the prevalent *Dc* because the prevalent maximal patterns are compact representations of all prevalent patterns and can be used to derive all prevalent patterns. The gap between the number of prevalent maximal $Dc$ and that of the prevalent $Dc$ is large, and thus the difference in efficiency between mining the prevalent maximal $Dc$ by our method and mining the prevalent $Dc$ by the traditional method is obvious.

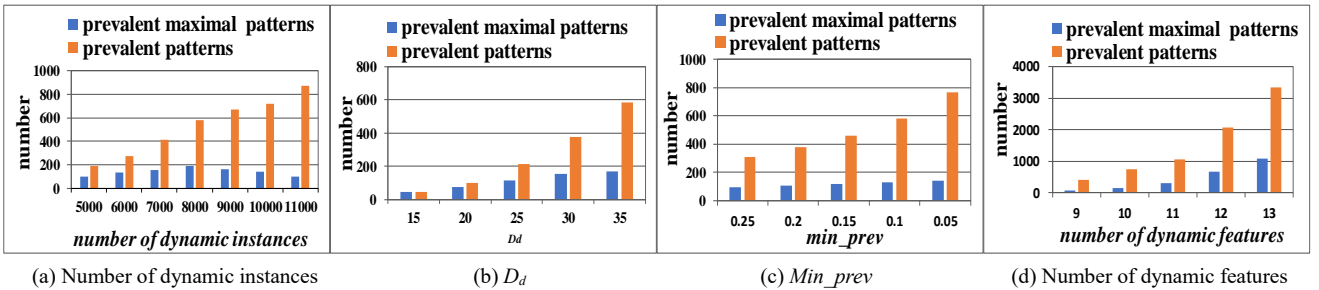

(a) Number of dynamic instances  (b) $D_d$  (c) *Min_prev*  (d) Number of dynamic features

Fig. 5. Representativeness of prevalent maximal *Dc*

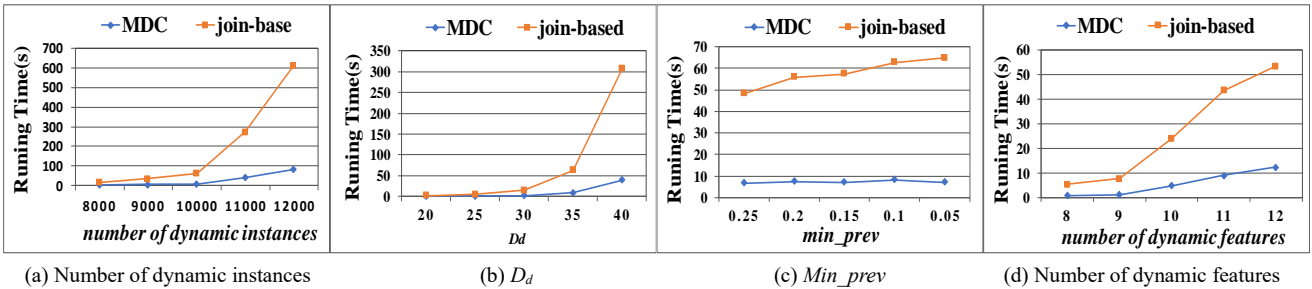

Fig. 6. Efficiency of Algorithm MDC

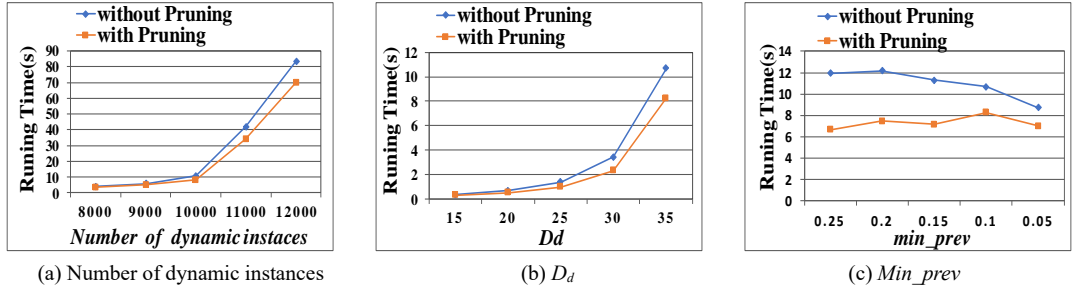

Fig. 7. Performance of pruning strategies

We propose an algorithm (i.e., Algorithm MDC) for mining the prevalent maximal $Dc$ to avoid the many connections and computations in the existing methods. We convert the size-2 prevalent patterns into a dynamic feature graph ($DG$) by *Algorithm1* such that the calculation and connection for the table instances are turned into the calculation and connection of the dynamic features. We obtain the set of maximal dynamic feature clique $Dfc$ from the dynamic feature graph $DG$ by *Algorithm2*. Finally, the maximal dynamic feature cliques as candidate maximal dynamic spatial co-location patterns are verified by *Algorithm3*, and we can obtain the prevalent maximal $Dc$. Moreover, we propose two pruning strategies to improve the efficiency of Algorithm MDC.

The experimental results from a real dataset and a synthetic dataset show that our algorithm can effectively mine the prevalent maximal $Dc$, that the number of prevalent maximal $Dc$ is much less than the number of all prevalent $Dc$ and that the performance of Algorithm MDC is better than the join-based [15,16].

The biggest limitation of our method is that parameters are designated by domain experts, such as $Dd$, *min_prev*, *time_span*, which is the common problem of mining spatial co-location pattern and is also the future research directions: 1) parameters might be learned from the dataset that can reduce the subjectivity of the parameter designated by experts as much as possible; 2) methods for how to set a more reasonable *time_span* can be considered to maximize the value/meaning of the dynamic spatial co-location patterns; 3) more efficient approaches to mining the prevalent maximal dynamic spatial co-location patterns can be designed. Our study also opens the door to exploring the dynamic relationships among spatial features.

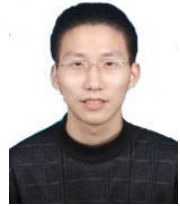

**Xin Hu** was born in Mianyang, China in 1988. He received the B.S. degree in computer science and technology from Capital Normal University, Beijing, China, in 2011, M.S degree in computer application technology from Yunnan University, Kunming, Yunnan, China, in 2014 and PhD degree in computer application technology from Beijing Normal University, Beijing, China, in 2019.

He is an assistant professor of Yangtze Normal University. His main research interests include knowledge graph, question answering, data mining, crowdsourcing.

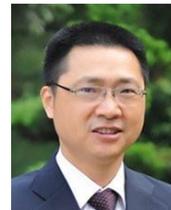

**Guoyin Wang (SM'03)** received the B.S., M.S., and Ph.D. degrees from Xi'an Jiaotong University, Xi'an, China, in 1992, 1994, and 1996, respectively. He was at the University of North Texas, and the University of Regina, Canada, as a visiting scholar during 1998-1999.

Since 1996, he has been at the Chongqing University of Posts and Telecommunications, where he is currently a professor, the director of the Chongqing Key Laboratory of Computational Intelligence, and the dean of the School of Graduate. He was appointed as the director of the Institute of Electronic Information Technology, Chongqing Institute of Green and Intelligent Technology, CAS, China, in 2011. He is the author of 10 books, the editor of dozens of proceedings of international and national conferences, and has more than 200 reviewed research publications. His research interests include rough set, granular computing, knowledge technology, data mining, neural network, and cognitive computing.

Prof. Wang is a senior member of the IEEE.

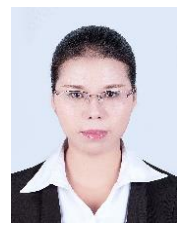

**Jiangli Duan** was born in 1989. She received the B.S. and M.S. degrees from Yunnan University, Kunming, Yunnan, China, in 2012 and 2017. She is currently pursuing the PhD degree in computer science and technology at Chongqing University of Posts and Telecommunications, Chongqing, China.

She is also an assistant professor of Yangtze Normal University, Chongqing, China. Her main research interests include knowledge graph, question answering, data mining, granular computing and cognitive computing.